\documentstyle[12pt]{article}

\font\sectionfont=cmbx12 at 12pt

\def\section#1{\vskip 1.5truepc\centerline{\hbox {{\sectionfont #1}}}
\vskip 1truepc\noindent\stepcounter{section}}
\font\authorfont=cmti12 at 14pt
\font\titlefont=cmbx12 at 16pt

\begin{document}
\setcounter{page}{1}

\def\sec#1{\vskip 1.5truepc\centerline{\hbox {{\sectionfont
#1}}}\vskip1truepc\noindent\noindent\stepcounter{section}}

\def\newsubsec#1#2{\vskip .3pc plus1pt minus 1pt\noindent {\bf #1}
{\rm #2}}
%%%%%%%%%  !!!!!!! %%%%%%%%%
% these commands  are used by  A.Losev
\newcommand{\be}{\begin{equation}}
\newcommand{\ee}{\end{equation}}
%%%%%%%%%%%%%%%%%%%%%

\newcommand{\eps}{\varepsilon}
\newcommand{\alp}{\alpha}
\newcommand{\gam}{\gamma}
\newcommand{\Aut}{{\rm Aut}\,}
\newcommand{\Out}{{\rm Out}\,}
\newcommand{\Mult}{{\rm Mult}\,}
\newcommand{\Inn}{{\rm Inn}\,}
\newcommand{\Irr}{{\rm Irr}\,}
\newcommand{\IBr}{{\rm IBr}}
\newcommand{\Ker}{{\rm Ker}\,}
\renewcommand{\Im}{{\rm Im}\,}
\newcommand{\Ind}{{\rm Ind}}
\newcommand{\diag}{{\rm diag}}
\newcommand{\soc}{{\rm soc}\,}
\newcommand{\End}{{\rm End}\,}
\newcommand{\sol}{{\rm sol}}
\newcommand{\Hom}{{\rm Hom}\,}
\newcommand{\rank}{{\rm rank}\,}
\newcommand{\Syl}{{\rm Syl}\,}
\newcommand{\Tr}{{\rm Tr}\,}
\newcommand{\tr}{{\rm tr}\,}
\newcommand{\Gal}{{\it Gal}\,}
\newcommand{\Spec}{{\rm Spec}\,}
\newcommand{\ad}{{\rm ad\,}}
\newcommand{\Sym}{{\rm Sym}\,}
\newcommand{\Char}{{\rm char}\,}
\newcommand{\pr}{{\rm pr}}
\newcommand{\Rad}{{\rm Rad}\,}
\newcommand{\abel}{{\rm abel}}
\newcommand{\codim}{{\rm codim}}
 \catcode`\@=11
\font\twelvemsb=msbm10 scaled 1200
\font\tenmsb=msbm10
\font\ninemsb=msbm7 scaled 1200%msbm9
\newfam\msbfam
\textfont\msbfam=\twelvemsb  \scriptfont\msbfam=\tenmsb
  \scriptscriptfont\msbfam=\ninemsb
\def\msb@{\hexnumber@\msbfam}
\def\Bbb{\relax\ifmmode\let\next\Bbb@\else
 \def\next{\errmessage{Use \string\Bbb\space only in math
mode}}\fi\next}
\def\Bbb@#1{{\Bbb@@{#1}}}
\def\Bbb@@#1{\fam\msbfam#1}
\catcode`\@=12

\renewcommand{\thefootnote}{\fnsymbol{footnote}}
%\footnotesep6.5pt

\hfill{hepth/9801179}

\hfill ITEP-TH-31/97

\hfill YCTP-P5-97

\centerline{\titlefont  "Hodge strings" and}
\centerline{\titlefont   elements  of  K.Saito's theory}
 \centerline{\titlefont of  the Primitive
form}

\vskip 2pc

\centerline{\authorfont  A.Losev \footnotemark[1]}
 \footnotetext[1]{This work  was supported by  RFFI  grant  96-01-01101,
 PYI grant  PHY9058501 ,  DOE grant  DE-FG02-92ER40704
and by grant 96-15-96455 for support of scientific schools}

\vskip 3pc

\centerline{To appear in the proceedings of the Taniguchi
Symposium}
\centerline{ Topological field theories, Primitive forms
and related topics,}
\centerline{ Kyoto, December 1996}

\vskip 3pc

\centerline{\sectionfont  Abstract}
\vskip 8pt
\rm  The  "Hodge strings" construction
 of solutions to associativity equations is proposed.
From  the topological  string theory  point of view  this
construction
 formalizes  the
"integration over the position of the marked point" procedure for
computation
of amplitudes.  From  the mathematical point of view  the "Hodge
strings"
construction  is  just a composition  of  elements of  harmonic
theory
(known among physicists  as a $t$-part of  $t-t^*$  equations)  and
the K.Saito construction  of flat coordinates (starting from  flat
connection
with a spectral parameter).

 We  also show  how  elements of K.Saito theory of primitive form
 appear naturally
in the "Landau-Ginzburg"  version  of  harmonic theory if we  consider
the  holomorphic  pieces of  germs of  harmonic forms  at  the
singularity.

\section{1. Introduction  and summary}
The first  aim of this article is to explain
how and why some version of the  Hodge
(harmonic)  theory leads to
the specific map from  tensor powers of the vector space
to the cohomologies of the Deligne-Mumford compactification of
the moduli space of rational curves with $n$ marked points.
 In physics
such a map is called "generalized amplitudes in topological strings";
in mathematics, a particular case of this map is called
"Gromov-Witten" invariants.

The second aim is to  explain how  elements  of  the  K.Saito
theory of  Primitive form (that  ,among other things,
 provides solutions to
associativity ( WDVV)  equation
 \cite{BV})
naturally  arise from the
"Landau-Ginzburg"   version  of  Hodge ( harmonic)  theory.

In section 2, we remind that  "generalized  amplitudes"
in genus zero  (in  a  "topological  string without  gravitational
descendents")  are in a one-to-one
correspondence with the solutions to the WDVV  equation(4),
 thus our construction (called "Hodge strings")
  should   end   with the solution to this equation.

Sections 3 and 4 contain  motivations coming   from the   topological
string theory  for the
construction (they partly answer the question "why?") and
could be omitted by a reader who is a mathematician. In section 3, we review
the concept and structures of "conformal topological strings",
and in section 4 we describe the "integration over the position
of the marked point" procedure of computation of the
"amplitudes" in genus zero.  Along these lines we explain the origin of the
$QG_{-}$-system, that  contains
$Z_2$ graded vector space $H$, odd operators $Q$ and $G_{-}$,
even commuting $Q$-closed operators $\Phi_i$
(all operators are acting on $H$)   and  the bilinear
pairing $<>$ on $H$. The "integration
over the marked point" procedure shows what kind of structure
should we expect to see.

Section 5 is axiomatic: here we introduce the notion and general
properties  of  an abstract
$QG_{-}$ system.
 Then we show that  starting  with
the
$QG_{-}$ -system
$(H,Q,G_{-},\Phi_i,<>)$  having   Hodge property, Pairing  of the
cohomology property  and the Primitive element property
one can  canonically   construct a  solution to the WDVV equation.
The  construction is done  in two steps.
First, by comparing  two flat connections
(the"Hodge" connection  and the "Gauss-Manin"  connection)
on the bundle of  $Q(t)+zG_{-}$ cohomologies,
we  show the existence of a flat connection
with the  spectral parameter ($\nabla^{H}-z^{-1} C$),
which  is known in  physics as  the  $t$-part of
$t-t^*$  equations \cite{CV}.
Then, using  the Primitive element property,
we  construct a solution  to the WDVV equations,
like  K.Saito  did  in the theory  of Primitive  form.
This section answers
the question "how" and formally is independent of the previous
sections.
 Nevertheless, we try to comment "why" the construction goes
this way by referring to section 4.

 In section 6, we review the
so called "Landau-Ginzburg"  realization of the "Hodge strings" input
(in physics such a system is  known as  $N=2$  supersymmetric
Landau-Ginzburg  quantum mechanics).

Then,in section 7, we start out
by briefly  reviewing  (in subsection 7.1)
elements  of  K.Saito's  theory  of  primitive
form  in  the form  of   "good  section" and  in terms  of
$QG_{-}$ systems.
( In the Appendix  we  relate it to the
original formulation).
  To  reach  K.Saito's theory of  Primitive
form  from the
"Landau-Ginzburg"  system (for quasihomogeneous case)
 we  first  pass from the smooth  quickly vanishing forms
of the "Landau-Ginzburg" system   to the non-holomorphic
germs  of  forms  at a  singularity,  and then
take holomorphic pieces  of   germs.
 We find  that  holomorphic  pieces  of  germs coming  from the development of
harmonic forms of the "Landau-Ginzburg" theory
 satisfy two of K.Saito's conditions for a "good  section"
and,
with  a quasihomogeneous "antiholomorphic superpotential
$\bar{U}$ ",
satisfy the third condition
 (this  third condition  is not necessary
for construction  of the solution to the WDVV equations) .

We explain  the  ambiguity of the
solutions  for  K.Saito's  conditions
for a  "good  section"   as coming
from  the "antiholomorphic superpotential  $\bar{U}$"
that disappears
 in the "taking  holomorphic pieces"
procedure  (this phenomena  in  topological
strings is called the "holomorphic anomaly").

We expect that methods developed here could
be useful in the understanding of  non-quasihomogeneous
systems.

{\bf   Conventions}.
The sum over  repeating indexes
(the physicist's  convention)  is adopted
in  the text .

\section{2. "Compact"  topological strings and the associativity equation}

"Topological string theory"\cite{Wi, DW,VV,DVV,Wi2,KM,BCOV}
studies  genus $q$
"generalized amplitudes" $GA_{q}$, taking values in cohomologies of
the
Deligne-Mumford compactification $\bar{M}_{q,n}$ of the moduli space
of complex structures of genus $q$ Riemann surfaces with $n$ marked
points.
Pairing between  $GA_q$ and the cycle $C\in \bar{M}_{q,n}$ is
given by the functional integral
\cite{Wi,DW,KM}
\be
(GA_q,C)(V_{1}, \ldots, V_{n})=
\int_{C \in \bar{M}_{q,n}} \int {\cal D}\phi
V_{1}(\phi(z_1)) \ldots
V_{n}(\phi(z_n))  \exp(S_{TS}(\phi)),
\ee
fields $V_i(\phi(z))$ are called "vertex operators" and
ordinary "amplitudes"
$A_q(V_1,\ldots, V_n)$ correspond  to
$C=\bar{M}_{q,n}$.

Deligne-Mumford compactification $\bar{M}_{0,n}$ is a union of
$M_{0,n}$ (a set of $n$ noncoincident points on $CP_1$ moduli $SL(2,C)$
action) and the compactification divisor $Comp$.The divisor $Comp$ is a
union
of components $C(S)$, where $S$  partitions  $n$ marked points
into  two groups  consisting of $n_1(S)$ and $n_2(S)$ points,
$n_i>1$. A surface corresponding to a general point in $C(S)$ is a
union of two spheres having one common point with $n_1(S)$ marked
points on the first sphere and $n_2(S)$ on the second. The set
of general points in $C(S)$ form the space
$M_{0,n_1+1} \otimes M_{0,n_2+1}$.

In this  paper,  we will  consider  a  class  of   "compact"
topological
string  theories that  have no gravitational descendents \cite{Wi}  among
its  "vertex  operators"  and  have
a nondegenerate pairing  on the space  of  "vertex  operators".
This class of theories includes, for example,  topological
sigma models of type A  on compact  Kahler  manifolds
and  twisted  unitary superconformal  theories.
It is believed  that  these  theories  play the same role among
all theories as smooth compact  manifolds  among  all manifolds.
It  expected that,   in  "compact"  topological  theories,
the functional integral for surfaces corresponding
 to points in $C(S)$ factorizes and \cite{Wi}
\begin{eqnarray}
&&(GA_{0},C(S))(V_{i_1},\ldots,V_{i_n})= \nonumber \\
&&  \eta^{jk}
 A_0(V_{i_1},\ldots,V_{i_{n_{1}}},V_j)
 A_0(V_{i_{n_1+1}},\ldots,V_{i_{n_2+n_1}},V_k)
\end{eqnarray}
where
$\eta$ is a matrix of symmetric bilinear nondegenerate products on
 vertex operators.

Keel  \cite{Ke} found that the homologue ring $H_*$ of $\bar{M}_{0,k}$
 is generated
by cycles $C(S)$.He described relations between these cycles
in homologies
leading
 to constraints on $GA_0$ because of (2).

An elegant way of formulating these constraints
uses the generating function for "amplitudes".
Introducing formal parameters $T_i$, we define  the germ  $F(T)$:
\be
F(T)=\sum_{k=3}^{\infty} \frac{1}{k!} A_0( T_{i_1} V_{i_1},
\ldots, T_{i_k} V_{i_k})
\ee
Then, Keel's relations lead to:
\be
\frac{\partial^3 F(T)}{\partial T_i \partial T_j \partial T_k}
\eta^{kl}
\frac{\partial^3 F(T)}{\partial T_l \partial T_p \partial T_q}=
\frac{\partial^3 F(T)}{\partial T_i \partial T_p \partial T_k}
\eta^{kl}
\frac{\partial^3 F(T)}{\partial T_l \partial T_j \partial T_q}
\ee
Using the factorization property and Keel's description of
homologies of  moduli space, we can reconstruct $GA_0$
from $A_0$ \cite{KM}, see also \cite{DVV}.

\section{3. Amplitudes in  conformal topological strings theory}
The "Hodge string" construction generalizes the
"integration over the position of the marked point " procedure
\cite{VV,DVV,Lo1,LP,BCOV, Lo2} of computation of amplitudes
in "conformal topological theory
 coupled to topological gravity" also known as
"conformal  topological string theory".

The general covariant action $S_m$ of topological field theory
is a sum of a "topological"(metric independent)
$Q$-closed term $S_{top}$ and
a $Q$-exact term for a fermionic scalar symmetry $Q$:
$$S_m=S_{top}(\phi)+Q(R(\phi),g),$$
where $g$ denotes the metric on the Riemann surface.
The energy-momentum
tensor $T$ is $Q$-exact:
\be
T=Q(\frac{\delta R}{\delta g})=Q(G)
\ee
We call topological field theory conformal, if $R$ is conformally
invariant, i.e. $G$ is traceless.

 We introduce fermionic two-tensor fields $\psi$,
such that functions of $g$,$\psi$ are forms on the space of
metrics. An external differential on these forms could be written as
follows:
$ Q_g=\psi \frac{\delta }{\delta g}$.

The action for a topological theory coupled to topological gravity
is
 $$S_{TS}=S_m+\psi G= S_{top}+ (Q+Q_g)(R).$$
The functional integral $Z(g,\psi)$ over the set of fields
 $\phi$ with the action
$S_{TS}$ is a closed form
on the space of metrics.
Since $G$ is traceless, $Z$ is a horizontal \cite{DVV,Di}
\footnote{Differential form on the principal
bundle is called horizontal if its contraction with the vertical
(tangent to fiber) vector is zero.Closed horizontal forms on the
total space correspond to closed forms on the base of the bundle}
 form with
respect to the action of conformal transformations of metrics
and diffeomorphisms of the Riemann surface; thus, it defines a closed
form on the moduli space of conformal(=complex) structures on the
genus $q$ Riemann surface.

To construct generalized amplitudes we insert
fields (zero-observables ="vertex operators")
 $V_{i}$ at marked points
on
Riemann surface. They should satisfy
\be
Q(V_i)=0,
G_{0,-}(V_i)=0.
\ee
Here $G_{0,-}$ is the superpartner of the component of the
energy-momentum
tensor $T_{0,-}$ that corresponds to the rotation with the constant
phase
 $z \rightarrow e^{i\theta}z$ of the local coordinate at the marked
point.
The first condition in (6) is needed to construct a closed form on the
space
of metrics, while the second provides horizontality of the
corresponding form with respect to diffeomorphisms that leave marked
points
fixed but rotate local coordinates \cite{Al,DN,DVV,Eg,Di}.

\section{4.  Integration over positions of marked points}
The "integration over marked points" procedure reduces  all
genus zero amplitudes to the three point amplitude:
$$F_{ijk}=A_0(V_i,V_j,V_k),$$ which can be computed from
topological
matter theory.

In conformal topological theory, we associate  a
two-observable $V_{i}^{(2)} = G_{L,-1}G_{R,-1}V_i$ to a zero observable
$V_i$.
Thus, we deform topological theory to a family
of theories parametrized by $t$, with the action
$S_m(t)=S_m+t_i V_{i}^{(2)};$
thus, zero-observables $V$ form a tangent bundle to this space
 of theories \cite{DVV}.

If, in the functional integral that computes the n-point  amplitude,
 we first   pick up one of the marked points
(we will call  it a  "moving point"),  integrate over the position
of the
moving
point, and only then take the functional integral, the
 $n$-point amplitude becomes the derivative
in $t$ of the $n-1$ point amplitude.

In the process of integration, we should take special care about
the region where the moving point tends to hit a fixed point
because the geometry there is not a naive one.
The contribution from this region(contact terms \cite{VV,Lo1,Lo2,LP,Di,BCOV})
leads to
a specific contact term connection on the bundle of zero-observables
over the
space of theories and thus on the tangent space to the space of
theories.

Repeating this procedure again and again, we can recover amplitudes
from
$F_{ijk}(t)$.
The amplitudes should be symmetric and independent of the order of
integration over positions of marked points.

In other terms, generating parameters $T$ from (3) should become
the so-called special coordinates on the space of theories,
the derivatives with
respect to the special coordinates should become covariantly constant
sections
of the contact term connection, and symmetric tensor $F_{ijk}$
(in the special coordinate frame) should be a third derivative of $F(T)$.
Moreover, $F(T)$ has to solve the WDVV equations (4).

All this implies that the contact term connection is quite a special
one!

To gain better understanding of this connection,
we will study the space of states in 2d theory associated with the
 circle (considered as a  component of  the boundary of  the Riemann
surface).
Moreover, we will restrict ourselves to the subspace $H$ of these
states
that are invariant under constant rotation of the circle.

Fermionic symmetry $Q$ of the theory and $G_{0,-}$
 reduce to  odd anticommuting operators $Q$ and $G_{-}$ on $H$.

 Zero-observables $V_i$
 (being inserted at the middle of the punctured disc)  generate
states $h_i$ that are $Q$ and $G_{-}$ closed:
\be
Q h_i= G_{-} h_i =0,
\ee
the zero observable $1$ generates the distinguished state $h_0$.
The operation of sewing two discs together corresponds to the
bilinear
pairing $<,>$.
Integrals of zero observables
along the boundary give operators
$\Phi_i=\int_{S_1} V_{i}d\sigma$.

One can show from the functional integral that
the objects defined above  have the following properties:
\begin{eqnarray}
&& Q^2=G_{-}^2=QG_{-}+G_{-}Q=0,
[Q,\Phi_i]=0,[\Phi_i,\Phi_j]=0,\\
&& Q^T=E Q, G^T=-E G,
\Phi^T= \Phi
\end{eqnarray}
Here transposition "T" is taken with respect to the
pairing $<,>$, and operator $E$ commutes with
$\Phi$ and anticommutes with $Q$ and $G_{-}$.

In the deformed theory,
$Q(t)=Q+[G_{-},t_i \Phi_i]$  in  the first order in $t$. To
ensure it globally we will take for
 simplicity\footnote{in general case one has to go in for
Kodaira-Spencer type arguments,see \cite{BCOV}}
\be
[[G_{-},\Phi_i],\Phi_j]=0.
\ee

The contribution from the region near the place where the
"moving" i-th point hits the marked j-th one gives the
"cancelled propagator argument"(CPA) connection
on states $h_j$ over the space of theories \cite{VV,Di,LP,Lo2}:
\be
\delta_{i}^{(CPA)} h_j = G_{-} \int_{0}^{\infty} d\tau G_{0,+}
\exp(-\tau
T_{0,+}) \Phi_i h_j,
 \ee
 thus $\delta^{(CPA)}h$ is $G_{-}$-exact.  Here $T_{0,+}$ is the
Hamiltonian acting on the space $H$, and $G_{0,+}$ is its
superpartner:
$T_{0,+}=Q(G_{0,+})$.

Covariantly constant
 sections\footnote{Flatness of CPA connection is necessary
for the consistency of the procedure}
 of the CPA connection  will be denoted as
$h_i(t)$.
This connection induces the connection on the space of
 zero-observables: covariantly constant sections of contact term
connection
$$V_i(t)=u_{i}^{j}(t) V_j$$ are such that, being inserted
 in the middle of the disc
in the $t$-deformed theory, they produce covariantly constant sections
$h_i(t)$:
\be
h_i(t)=lim_{r \rightarrow 0}
 u_{i}^{j}(t)
 r^{T_{0,+}} \Phi_j h_0(t).
\ee
Let us denote as $C_i(t)$ the linear operator representing
the action of $\Phi_i$ in
$Q(t)$-cohomologies. Then, the relation (12) reads:
\be
[h_i(t)]_{Q(t)} = u_{i}^{j}(t) C_{j}(t) [h_0(t)]_{Q(t)}
\ee
here and below  $[h]_{Q}$ stands for a class of a $Q$-closed
element $h$ in $Q$-cohomologies.

 From the functional integral we
get:
 \be
 F_{ijk}(t)=<h_i(t),\Phi_{l} h_k(t)> u_{j}^{l}(t).
 \ee

While the string origin of the described procedure is quite natural,
its consistency is far from being obvious.

 \section{5. The  "Hodge string"   $QG_{-}$-system}
\newsubsec{5.1  General facts about  $QG_{-}$  systems}\\
\noindent {\bf   Definition.}  The   $QG_{-}$ system
$ (Q,G_{-}, \Phi,  H)$   is
a collection of
 $Z_2$ -graded vector space $H$, odd operators $Q$ and $G_{-}$ ,
and a  set of  even operators $\Phi_i$,  $i=1,\ldots, \mu$,
acting on this space, that
have the properties (8,10).

Given  a  $QG_{-}$ system,  one can construct
a family  $Q(t)$  of nilpotent  odd operators
 in $H$:
\begin{equation}
Q(t)= Q+ t_i [ G_{-}, \Phi_i ]
\end{equation}
over  a  deformation space  with coordinates $t_i$.

\noindent{\bf  Definition. } {\it  Cohomologies of  $QG_{-}$-systems.}\\
Let   $H_{Q(t)}$  be the  space  of  $Q(t)$ cohomologies
in  $H$.
Let
\begin{equation}
Q(t,z)=Q(t) + z G_{-}
\end{equation}
\begin{itemize}
\item  Let  $H_{Q(t)}$ be the space of cohomologies of  $Q(t)$ in $H$
\item  Let  $\hat{H}_{Q(t,z)}$  be the space
of  cohomologies of  $Q(t,z)$  in  $H\otimes C[[z]]$
\item  Let  $H_{Q(t,z)}$  be the  space of
cohomologies of  $Q(t,z)$ in the space
$H  \otimes C<< z>>$,
where  $C<<z>>$ is the space of Laurent
expansions in $z$
\item  Let  $H_{Q(t,z)}^{l}  \subset  \hat{H}_{Q(t,z)}$  be the  space
 of  "little" cohomologies,
defined as  those classes  in   $\hat{H}_{Q(t,z)}$ that have
representatives in  $H$:
\begin{equation}
H_{Q(t,z)}^{l}= \{  [\omega]_{Q(t,z)} \in  \hat{H}_{Q(t,z)} |
\omega \in H,  Q \omega = G_{-} \omega=0 \}
\end{equation}
\end{itemize}

\noindent{\bf   Remark.}The space $\hat{H}_{Q(t,z)}$ has a natural
decreasing filtration
by powers of $z$:
\begin{equation}
\hat{H}_{Q(t,z)} = \hat{H}^{(0)} \supset  \hat{H}^{(1)} \supset  \ldots
\end{equation}
a  class is in  $\hat{H}^{(k)}$  if it  contains element  $z^k\omega$.
The inclusion  $H_{Q(t,z)}^{l}  \subset  \hat{H}_{Q(t,z)}$ induces the
decreasing  filtration on "little" cohomologies.

\noindent{\bf   Remark from  string theory.}
In string theory of the general type
the space of  "little" cohomologies
corresponds to the space  of states created  by
"vertex operators".  One can show  \cite{VV,Lo1, Eg, Lo2}
that  states  from  $H_{Q(t,z)}^{l,(k)}$ are created  by  "vertex
operators" that  are $k$-th gravitational descendents.

\noindent{\bf   Definition.} Let  $C_i(t)$ : $H_{Q(t)} \rightarrow  H_{Q(t)}$
 be a linear operator representing the  action of $\Phi_i$
 in  $Q(t)$ cohomologies:
\begin{equation}
C_i(t) [ \omega ]_{Q(t)} = [ \Phi_i  \omega ]_{Q(t)}
\end{equation}
\noindent{\bf Remark.} Operators $C_i(t)$ should be considered
as components of the one-form on the deformation space with values
in $End H_{Q(t)}$. From the definition it follows that these operators
commute with each other:
\begin{equation}
[ C_i(t), C_j(t) ] =0
\end{equation}
 \noindent {\bf   Definition.} A  morphism  of $QG_{-}$-systems
\begin{equation}
(Q^{1}, G_{-}^{1}, \Phi_{i}^{1}, H^{1} ) \rightarrow
 (Q^{2}, G_{-}^{2}, \Phi_{i}^{2},  H^{2} )
\end{equation}
is  a  morphism  $H^{1} \rightarrow  H^{2}$
commuting with the action of operators
$Q^{\beta}, G_{-}^{\beta}, \Phi_{i}^{\beta} $
in  $H^{\beta}$, $\beta=1,2$.

It is clear that the morphism of  $QG_{-}$ systems induces
the morphism of  cohomologies  of $QG_{-}$ systems.

\noindent {\bf   Definition.} A morphism  of $QG_{-}$-systems will
be called a  {\it  quasiisomorphism}  of $QG_{-}$ systems
if  it induces  an isomorphism  in  {\it   all}  cohomologies
of  $QG_{-}$ systems.

\noindent {\bf  Definition.} By the "Gauss-Manin" connection
in a $QG_{-}$ system, we  call  a   canonical  flat  connection
$\nabla^{GM}$ in
$H_{Q(t,z)}$
over  $C[[t]]$, whose  horizontal sections
$[\omega^{GM}(t,z)]_{Q(t,z)}$
satisfy the following:
\begin{equation}
[\omega^{GM}(t,z)]_{Q(t,z)}=[ \exp(- t_i \Phi_i/z)
\omega^{GM}(0,z)]_{Q(t,z),}
\end{equation}
i.e. their   representatives  solve the following differential equation:
\begin{equation}
\frac{\partial}{\partial  t_i}  \omega^{GM}(t) +z^{-1} \Phi_i
\omega^{GM}(t)   \in   Im (Q(t,z))
\end{equation}

\noindent {\bf  Remark.}  We call this canonical connection  "Gauss-Manin"
following  K.Saito  (see the Appendix).

\noindent{\bf  Remark.} It is clear that  morphisms of $QG_{-}$ systems
induce morphisms  of the  "Gauss-Manin"  connections,
and  quasiisomorphisms induce the isomorphism of these
connections.

Below  we will list   some additional  properties  that the $QG_{-}$  system
could have
and that would be important for the "Hodge string"  system .

\noindent {\bf The Hodge property.}
 \begin{equation}
 Im Q \cap  Ker G_{-} =Im G_{-} \cap  Ker Q = Im(QG_{-})
\end{equation}

\noindent{\bf  Statement  5.1}  It follows from the Hodge property that
$dim H_{Q}= dim H_{G_{-}}$ and there  exists a set  $\{  h_a \}$ of
$Q$  and $G_{-}$ closed
 elements  of  $H$ (these  elements are unique  up to  $Im QG_{-}$), such that
 classes  $[h_a]_Q$ and  $[h_a]_{G_{-}}$ form
bases in  $Q$  and  $G_{-}$ cohomologies.

\noindent{\bf  Remark.}  In harmonic theory such  elements are  just harmonic
forms;
that is  why below we  will call  these elements
"harmonic" .

\noindent{\bf   Statement  5.2}
If the $QG_{-}$  system  has  a  Hodge property, then
\begin{equation}
H_{Q(0,z)}  \cong  \hat{H}_{Q(0,z)}
 \otimes C[z^{-1}]  \cong H_{Q(0)}\otimes C<< z>>
\end{equation}
and
\begin{equation}
H_{Q(0,z)}^{l} \cong H_{Q(0)}.
\end{equation}

\noindent{\bf  Proof.}
Classes in the first  line are identified  by considering
  $h P(z) $  and  $h P(z,z^{-1})$
for  "harmonic"  $h$  as  representatives of
classes in    $\hat{H}_{Q(0,z)}$  and   $H_{Q(0,z)}$  respectively.
(Here,  $P$ are polynomials. )
The statement on the second line becomes clear  as  a generalization
of the following reasoning: $[\omega_1] =z [\omega_2]$, if
$\omega_1 = Q \omega'$  and  $\omega_2= - G_{-} \omega'$.
But   from the Hodge property it follows  that
$\omega_i  \in Im QG_{-}$, and  thus  $\omega_i  \in Im (Q+zG_{-})$.
$\Box$

\noindent {\bf  Remark.}
The difference between $H_{Q(0,z)}^{l}$  and  $H_{Q(0)}$ is one
of the criteria  showing the failure of the Hodge property.
From the string theory point of  view  this difference  means
that  corresponding  string theory  has gravitational  descendents
among its vertex operators, and is "noncompact" in  the sense
of section 2.

The $QG_{-}$ system with pairing is a
 $QG_{-}$ system  with the  bilinear  pairing   $<,>$
satisfying  property (9).

\noindent {\bf  Remark.}From  (9) it follows that the pairing
$<,>$  descends to both  $Q(t)$
 and  $G_{-}$  cohomologies,
i.e.
 for  $Q(t)$-closed  $\omega_2  \in H $
 \begin{equation}
<Q(t) \omega_1,  \omega_2>=< \omega_2, Q(t) \omega_1>=0,
\end{equation}
and  for  $G_{-}$ closed $\omega_2$
\begin{equation}
<G_{-} \omega_1,  \omega_2>=< \omega_2, G_{-} \omega_1>=0,
\end{equation}
\noindent {\bf   Pairing  of the cohomology property.}\\
The pairing  $<,>$ is non-degenerate when restricted
to $Q$-cohomologies.

\noindent {\bf Primitive element property.}
$Dim H_{Q(0)} =\mu  $  and  there is a class $[h_0]_{Q}$ in $Q$-cohomologies
(that we  will call  a primitive class)
such that  the set  $\{  C_i([h_0]) , i=1,\ldots,\mu \}$
forms a basis  in $H_{Q(0)}$.

\noindent{\bf  Remark.} The  primitive element  is not  unique.
One can easily see that if the $QG_{-}$ system has a
primitive class, almost all classes  are primitive\footnote{The work
\cite{Kr}  implicitly assumes that  it is possible  to construct
solutions  to WDVV equation starting from  any primitive element}.

\newsubsec{5.2  Solution  to the WDVV equations from the
"Hodge string"  $QG_{-}$-system.}\\

\noindent{\bf   Definition.}
The  "Hodge  string" system is a $QG_{-}$ -system with
pairing that  has the Hodge property,
pairing of cohomologies property and the primitive element
property.

\noindent{\bf   Theorem.}
There is a  canonical construction of  a solution to the WDVV equation
from  the  "Hodge string" system  and  the
choice  of   a  Primitive  element. \\

\noindent{\bf   Summary  of the construction.}
The construction is made in two steps. In the first step, we
start  from the  "Hodge string"  system  and
construct a flat connection with the spectral parameter .
Physicists know  this connection
as the $t$-part of the $t-t^*$ equations \cite{CV}.
The Primitive element property
 is not used in the first step.

The flat connection with the spectral parameter  appears  from
the comparison  of two flat connections.
The first connection would be the "Gauss-Manin" connection.
The  second   connection  comes as a formalization of the
CPA-connection
on the space of states (see section 4) -
variation of  its covariantly constant  section will be
$G_{-}$ exact.
Since  such a connection exists canonically  due to
the "Hodge"  property,  it will  be called below the  "Hodge"
connection .
This connection could be extended to connection in
$Q(t,z)$ cohomologies, and we will  call this extension
 Hodge connection too.
The difference between these two connections turns out to be
one-form $C_i(t)$ (introduced in subsection (5.1)) divided by $z$
(that becomes a spectral parameter).

 In the second step, with the help of the Primitive
element property, we induce a  flat connection on the tangent bundle
to the deformation space from the  "Hodge" connection constructed in step
one
(i.e., we induce a connection on the space of zero-observables from the
connection on the space of states, as in (13)). Then, we integrate
covariantly constant vector fields of this connection to special
coordinates
$T$ on the deformation space and, finally, construct   $F(T)$.
This step was first done by K.Saito (for a  $QG_{-}$ system  coming
from
the family of hypersurfaces near the singularity).

\noindent {\bf    Construction.}

\noindent{\bf  Statement. Step 1.} The "Hodge String"  system leads
canonically to
the matrix-valued  1-form   on the deformation space $C_{i,ab}(t)$,
 (first constructed by K.Saito \cite{Sa} in a
slightly different context) such  that  the following differential
operators commute
 \be
 \frac{\partial}{\partial t_i} \delta_{ab} - z^{-1} C_{i,ab}(t),
\ee
 and   $C_{i,ab}=C_{i,ba}$.

\noindent {\bf  Proof  of Step 1 .}
The proof of the Step 1  is divided  in two parts, 1A  and 1B.

\noindent{\bf  1A. Hodge  connection in  $H_{Q(t)} $:}
The idea  of construction  of the "Hodge"  connection is as follows.
The Hodge property  canonically  identifies  $Q(0)$- cohomologies
and  $G_{-}$-cohomologies.  While  the  operator $Q(t)$  changes
with $t$,
the operator  $G_{-}$ remains fixed.  Since variation of  $Q(t)$  is
$G_{-}$-exact,
it is possible to  identify  canonically   $Q(t)$ and $G_{-}$
cohomologies  over $C[[t]]$,
and construct  such  a  flat connection  in  $Q(t)$  cohomologies
that  the  image in  $G_{-}$ cohomologies of its covariantly constant sections
(that are taken to be $G_{-}$ closed)
is constant.  We will call  such a connection  the "Hodge" connection.

Specifically, let us define  $\Phi= \frac{t_i}{\tau}  \Phi_i$, and
consider   the following  equation:
\begin{equation}
(Q(0)+\tau [  G_{-}, \Phi  ] ) h_a(t)=0, \; h_a(t)=  h_a  +
\sum_{k=1}^{\infty}\tau^{k}
\omega_k
\end{equation}
with  $h_a$ being the $Q(0)$ and $G_{-}$ - closed element of  $H$.
This  leads to
\begin{equation}
Q \omega_1 = - G_{-} \Phi  h_a,
Q\omega_{k+1} =-G_{-} \Phi   \omega_{k}
\end{equation}
and we should like to solve these equations for $G_{-}$-exact $\omega_k$.

Due to the Hodge property, it is possible to solve recursively  the
equations above  for any $Q(0)$  and  $G_{-}$  closed  element
$h_a$.  The solution $h_a(t)$ is  a germ in $t$
defined  up  to  $Im  (Q(t)G_{-})$  (considered as the germ in  $t$).

The germ  $h_a(t)$ satisfies the  following  differential equation:
\begin{equation}
\frac{\partial}{\partial  t_i} h_a(t) - G_{-} \frac{1}{Q(t)}
(\Phi_i  h_a(t) - C_{i,a }^{b} h_b(t))   \in  Im Q(t)G_{-};
\end{equation}
here, $C_{i,a }^{b}$ are  matrix elements of
the operator $C_i(t)$  written in the basis
$\{  [h_a(t)]_{Q(t)}  \}$ in $H_{Q(t)}$ ,  $ \frac{1}{Q(t)} $  has to be
considered as a
germ  in $t$  and
operation  $\frac{1}{Q}$
is defined only on the  $Q$-closed  elements
of $H$  and  stands for taking some preimage of $Q$.
This ends  the construction of  "Hodge" connection.

\noindent{\bf  Statement.} The "Hodge" connection preserves
the  bilinear pairing  $<,>$ on $Q(t)$  cohomologies.

\noindent{\bf  Proof.} The bilinear pairing  descends  not only to
$Q(t)$  but also to  $G_{-}$ cohomologies. However, the
class of  $G_{-}$ cohomologies of  $h_a(t)$
does not depend on $t$.$\Box$

\noindent  {\bf  1B. Comparison of  "Gauss-Manin"  and "Hodge"
connections}

In order  to compare  two connections,  we will lift  the
"Hodge" connection to the
connection   $\nabla^H$   in the bundle of  $H_{Q(t,z)}$
cohomologies. The covariantly constant
sections  of the lifted connection  are taken to be
equal to   $[ h_a (t) P_a(z)]_{Q(t,z)}$,
where $P_a(z)$  is a  $t$ independent  element  of  $C<<z>>$.

Due to  the identity
\begin{equation}
G_{-} \frac{1}{Q(t)} \omega =-z^{-1}  \omega +
(Q(t)+zG_{-}) z^{-1} \frac{1}{ Q(t)}  \omega
\end{equation}
we find that
 the relation between
"Gauss-Manin" and  "Hodge" connections
 in $H_{Q(t,z)}$
takes the following form:
\begin{equation}
\nabla_{i}^{GM}=\nabla_{i}^{H} - z^{-1} C_{i}
\end{equation}
where  $C_i$ is the matrix representing the action of
$\Phi_i$  in $Q(t)$ cohomologies.

Now let us  rewrite the above  relation  in the basis
of covariantly constant sections of the "Hodge" connection:
\begin{equation}
\nabla_{i}^{GM}= \delta_{a}^{b} \frac{\partial}{\partial t_i} -
z^{-1}
C(t)_{i,a}^{b}
\end{equation}

Since  the bilinear  pairing  is  preserved  by the "Hodge" connection, it
is  represented by  a  $t$ - independent  bilinear form  in the
basis we are working with.  Moreover, the  bilinear form  is
non-degenerate
(due to a corresponding property) and
from the  property (9) we conclude that  $C_{i}^T=C_{i}$.
Making a change in the basis that  puts  the pairing into the form
$\delta_{ab}$,  we prove the assertion  made in the step 1.

\noindent {\bf Step 2.}
From  the assertion  in  Step 1 we conclude that
\begin{equation}
\frac{\partial}{\partial t_i} C_{j,ab}(t)=\frac{\partial}{\partial t_j}
C_{i,ab}(t)
\end{equation}
so  there exists a symmetric matrix $\tau_{ab}(t)$, such that
\begin{equation}C_{i,ab}(t)= \frac{\partial}{\partial t_i}\tau_{ab}(t).
\end{equation}

\noindent {\bf  Definition.}Fix  the Primitive element  $[h_0]_{Q(0,0)}$.
Let  $h_0$ be
the harmonic representative of   $[h_0]_{Q(0,0)}$. Let  $h_0(t)$ be the
result of the transport  of this element  by  the "Hodge" connection, so that
$h_0(t)= h_{0,b} h_b(t)$ . Here  $h_b(t)$ stands for the basis
in covariantly constant  sections of "Hodge" connection. Note, that
coefficients  $h_{0,b}$ are $t$-independent.
Let us define the {\it  auxiliary}
special coordinates $\theta_a$ on the deformation space as:
\be
\theta_{a}(t)=\tau_{ab}(t) h_{0,b}.
\ee

\noindent{\bf Statement.}There exists a function $F(\theta)$
of the auxiliary special coordinates  defined by
\be
\frac{\partial^2 F(\theta)}{\partial \theta_a \partial
\theta_b}=\tau_{ab}(t(\theta))
\ee
such that it satisfies the WDVV equations with
$\eta^{ab}=\delta^{ab}$.

\noindent{\bf Proof.} Explicit check.\\

\noindent{\bf   Definition.}
 We define the special coordinates\footnote{the coordinates
$T_i$ integrate vector fields $u_i$ introduced in (12,13)}
 $T_i$ as linear combinations
of  $\theta_a$ by:
\be
\theta_a=T_i C_{i,ab}(0)h_{0,b}.
\ee

\noindent{\bf   The   result  of the "Hodge string" construction }

The function $F(\theta_a(T_i))$ is the desired function that solves
the associativity
equations with $\eta^{ij}$,such that its inverse is given by:
\begin{equation}(\eta^{-1})_{ij}=<h_0, C_i(0) C_j(0) h_0>=h_{0,a}
(C_i(0) C_j(0))_{ab}
h_{0,b}\end{equation}

Below, we will present some explicit formulas. Let us define the
coefficients  $C_{i j_1\ldots j_n,ab}$  as
\be
C_{i,ab}(t)=\sum C_{i j_1\ldots j_n,ab} \frac{t_{j_1}\ldots
t_{j_n}}{n!},
\ee
Then, we have the following formulas for  the "amplitudes"
\begin{eqnarray}
&& A_0(V_i,V_j,V_k)=<h_0, C_i C_j C_k h_0>,\\ \nonumber
 &&A_0(V_i,V_j,V_k, V_l)=<h_0, C_i [C_j,C_{kl}] h_0>\\ \nonumber
&& A_0(V_i,V_j,V_k, V_l,V_m)=
<h_0, C_i [C_{jkl}, C_m] h_0>+ \\ \nonumber
&& <h_0 , [C_{im},C_{j}] C_{kl} h_0> +
  <h_0 , [C_{im},C_{l}] C_{jk} h_0> + <h_0 , [C_{im},C_{k}] C_{lj}
h_0>
\end{eqnarray}

\section{6. The "Landau-Ginzburg" realization of Hodge data}

Here, we present the realization of the "Hodge string"  system  coming
from the
$N=2$ supersymmetric quantum mechanics on $C^d$ with superpotential, see
\cite{CGP,Ce,CV} and references therein.

 Let us denote as
$X^A$  the  holomorphic coordinates on $C^d$ and let us take
two polynomials: a holomorphic  polynomial  $W(X)$ (in physics it is
called
superpotential) and an antiholomorphic\footnote{here and below
$\bar{X}$ denotes the complex conjugate of $X$} one  $\bar{U}(\bar{X})$.

Let the space $H^F$ be the space of smooth forms $\omega$ on $C^d$  :
\begin{equation}
\omega=
\omega(X,\bar{X})_{A_{1} \ldots A_{p}, \bar{A}_{1}  \ldots \bar{A}_q
}
dX^{A_1} \ldots  d\bar{X}^{\bar{A}_{q}}
\end{equation}
such that  any finite  number  of derivatives of  their coefficients
$\omega(X,\bar{X})_{A_{1} \ldots \bar{A}_q
}$
vanish
when $|X|\rightarrow \infty$ faster than any negative power of $|X|$
(F stands for  "fast  vanishing").
We take the odd operators $Q$ and $G_{-}$ to be :
\be
Q=\bar{\partial}+\partial W; G=\partial+ \bar{\partial}\bar{U},
\ee
here and below,
$\partial=dX^A \frac{\partial}{\partial X^A}$
and $\bar{\partial}$   is obtained from $\partial$
 by complex conjugation, $\partial W $ stands for  external
multiplication by the $(1,0)$ form   $dX^A \frac{\partial W}{\partial
X^A} $.

Let $\Phi_i=\Phi_i(X)$ be the set of polynomials
 that form a basis in the ring
$J(W)= C[X]/I(W)$, where the ideal $I(W)$ is generated by partial
derivatives
of $W$.

We will define pairing $<>$ between two forms from $H^F$ as:
\begin{equation} \label{pair}
<\omega_1, \omega_2>=\int_{C^d}\omega_1 {\cal C} \omega_2,
\end{equation}
where the Weil  operator
\begin{equation}
{\cal C}=(\sqrt{-1})^{(\hat{p}-\hat{q})},
\end{equation}
operators $\hat{p}=p, \hat{q}=q$
when acting on the $(p,q)$ forms. This pairing  has property (9)
with  $E=\sqrt{-1} (-1)^{(\hat{p}+\hat{q})}$.

We will study the properties of  the $QG_{-}$
system  defined above
with the help  of  some version of Harmonic theory.

Specifically, let us introduce  two auxiliary operators:
\begin{equation}
Q'(\bar{U})= *( \partial - \bar{\partial} \bar{U})*
\end{equation}
and
\begin{equation}
 G'_{-}(W)= *(\bar{ \partial } - \partial  W) *;
\end{equation}
here $*$  is a Hodge operation associated with the standard flat
Kahler
metric on $C^d$.

It is  easy  to check  that operators
$Q,Q',G_{-},G'_{-}$ have the same commutation relations as
$\bar{\partial}, \bar{\partial}^{+}, \partial ,\partial^{+} $
on a compact Kahler manifold with the
standard Laplacian $\Delta$  being replaced by
operator  $\Delta(W,\bar{U})$,
namely:
\begin{equation}
\{  Q(W),  Q'(\bar{U})  \}=\{  G'_{-}(W),  G_{-}(\bar{U})
\}=\Delta(W,\bar{U})
\end{equation}
\begin{equation}
\Delta(W,\bar{U})=\Delta -  \delta_{B \bar{B}} ( \frac{\partial
\bar{U}}{\partial \bar{X}^{\bar{B}}}
\frac{\partial W}{\partial  X^{B}}
+ \frac{\partial^2 W}{ \partial X^A \partial X^B} dX^A
\iota_{\frac{\partial}{\partial \bar{X}^{\bar{B}}}}+
 \frac{\partial^2 \bar{U}}{\partial \bar{X}^{\bar{A}}
\partial \bar{X}^{\bar{B}}} d\bar{X}^{\bar{A}}
 \iota_{\frac{\partial}{\partial X^B}}).
\end{equation}
Here,  $\iota_{v}$ denotes the operator of contraction of the form
with the vector field $v$.

We will begin with the case  $\bar{U}=(W)^{*}$.

\noindent{\bf  Statement  6.1 }
The harmonic forms (forms from the $Ker \Delta(W,(W)^{*})$)
form  bases in  spaces of
$Q(W),Q'(W^{*}),G_{-}(W^{*}),G'_{-}(W)$ cohomologies,
and   a pair  $Q$ and $G_{-}$ has the Hodge property.

\noindent {\bf  Idea of the proof.}  Operator $\Delta(W,(W)^{*})$ is
Hermitean  and has a discrete  spectrum; thus,  it is
a  complete analogue  of the ordinary Laplasian on
Kahler  manifolds, so the same proofs  are valid.$\Box$

Now let us examine the structure of cohomologies of
$Q(W)$.

\noindent{\bf   Statement  6.2}
The operator  $Q(W)$  has cohomologies
only in the middle dimension, and the
dimension of the space of these cohomologies
is equal  to  the dimension of the ring $J(W)$.

\noindent{\bf  Sketch of the proof.}
1) $Q$ has no cohomologies below the middle
dimension. Proof. Cohomologies of
the operator  of multiplication  by $\partial W$
 are non zero only in  $(d,k)$ components of the space
of forms. Thus , 1) follows from the spectral sequence argument.\\
2) $Q$ has no cohomologies above the middle dimension. Proof.
The Hodge $*$ operation identifies harmonic forms above the middle
dimension with those below the middle dimension(up to the change
of sign of $W$). Thus, 2) follows from 1)  and the Statement  6.1.\\
3) $Dim H_{Q} \geq  Dim J $. Proof.
Given a  polynomial  $\Phi_i$  representing  an element
of  $J$,  one easily constructs a form $\omega_i \in  H^F$
\begin{equation}
\omega_i = \exp ( -\{Q(W), R \} ) \Phi_i  dX^1 \ldots  dX^d
\end{equation}
where   operator  $R=*\bar{\partial}\bar{W}* $ .\\
4) There are no other cohomologies. Proof.
Consider the deformation of  $W$  into  a polynomial
having  only   simple critical points
widely separated from each other  (their number would be equal to
$dim J$).  Then,  from quasiclassical  arguments
one can show that  the number of harmonic forms
is not greater than the number of critical points.
The number of  harmonic forms  is invariant under
the deformation  due to 1),2)  since there is
an index ($Tr (-1)^{(p+q)} $) that is invariant under the
deformation.$\Box$

From the description of cohomologies it is obvious that  the
"Landau-Ginzburg"   $QG_{-}$ system has
the Primitive element property .

\noindent{\bf  Statement  6.3}
Pairing  $<>$ is nondegenerate.

\noindent{\bf  Proof.} Consider the value of pairing  on the forms
$\omega_i$ introduced  in the proof of  Statement 6.2.
After  some calculation we get:
\begin{equation}
<\omega_i, \omega_j>=
\sum_{\alpha} Res_{\alpha} \frac{\Phi_i  \Phi_j  dX^1 \ldots
dX^d}{\prod_{A=1}^{d} \frac{\partial W}{\partial  X^A}}.  \end{equation} here
sum is taken over all critical points  of  $W$, and  $Res_{\alpha}$  is the
residue at the critical point  $\alpha$.
$\Box$

Thus we found that  for  $U=W^*$
"Landau-Ginzburg" system  is
a "Hodge string"  system.

To see  the other options, we
will introduce the following definition:

\noindent{\bf  Definition.}
The polynomial  $\bar{U}$ is called  good  for  the polynomial $W$,
if the Hodge property  is satisfied.

\noindent{\bf  Statement  6.4}
If  the  operator  $\Delta(W,\bar{U}) $ acting on $H^F$ has a
discrete
spectrum, and
 $dim Ker \Delta(W,\bar{U})
= dim (H_{Q(W)})$,
   then  polynomial  $\bar{U}$
is  good for polynomial  $W$.

\noindent{\bf  Proof.}
For operators with a discrete spectrum the  eigenspaces with
non zero eigenvalues are  finite  dimensional   and
contain no cohomologies
(that is why  if  $\Delta(W,\bar{U}) $ has a discrete
spectrum,  $dim Ker \Delta(W,\bar{U})$  can not be smaller
than $dim (H_{Q(W)})$ or than  $dim (H_{G(\bar{U})}$ ) . These
eigenspaces  are  preserved  by  $Q$  and  $G_{-}$
and  the pair  $Q$   and $G_{-}$ restricted  to these
spaces has the  Hodge  property.
Thus, the pair  $Q$   and $G_{-}$  may  not
have  the Hodge property  only  when being restricted to an
eigenspace with a zero eigenvalue; however,   this space
 is just  annihilated  by $Q$  and  $G_{-}$ .$\Box$

\noindent{\bf Corollary 1 .}
Polynomial  $M^2 (W)^{*}$  is good  for a polynomial  $W$
for any real number  $M$.

\noindent{\bf   Proof.}
\begin{equation} \label{dil}
 \Delta(W,M^2 (W)^{*})= M^{-\hat{q}}  \Delta(M W,M (W)^{*})  M^{(\hat{q})};
\end{equation}
here, $\hat{q}$ stands for the  operator that acts on  $(p,q)$ forms
as  multiplication  by $q $.

\noindent{ \bf   Corollary  2.}
Consider  the space  $A$ of polynomials $\bar{U}$  such that
\begin{equation}
Re (\frac{\partial W}{\partial  X^{A}}
\frac{\partial \bar{U}}{\partial  \bar{X}^{A}} )
\rightarrow
+\infty
\end{equation}
as  $|X| \rightarrow +\infty$.
Then, there is  an open set  in this space consisting  of  polynomials
$\bar{U}$
that   are  good  for  $W$.

\noindent{\bf Sketch of the proof.}
For polynomials  from  $A$  the  operator $\Delta(W,\bar{U}) $  has a
discrete
spectrum ( forms  with  the bounded  real  parts of   eigenvalues
are  confined  in  $C^d$ by the growing potential  at infinity ; their
derivatives
are  also confined  due to the Laplasian).  The dimension of
the space  of harmonic forms for such   $\Delta(W,\bar{U}) $ can
increase
only  when  a  non zero  eigenvalue  comes down to zero;  this could
happen
only  outside an open set  containing  $\bar{U}=W^*$.

Corollary 2  shows that there are  many polynomials $\bar{U}$
that are  good for $W$,
and thus good  enough  to produce the "Hodge string"  system.

\vspace*{1cm}

\section{7. From  LG  harmonic theory  to  "good
section" of K.Saito}

\newsubsec{7.1  K.Saito  $QG_{-}$ system
 and conditions  for a "good section".}\\
We  will start  with a short  sketch  of   K.Saito  theory  of
primitive form
in the form of a  "good section"(\cite{Sa},  part 4)  in terms of
$QG_{-}$-systems.

Let   $W(X,t)=W(X)+t_i \Phi_i (X)$  be a versal deformation  of  the  isolated
singularity
$W(X)$    at    $X=0$ .
Let   $H^S$  be the  space  of germs  of   holomorphic
$(k,0)$  forms  at a  singularity .

K.Saito's theory  represents  operators of the $Q,G_{-}$ system  as
follows (K.Saito's representation is denoted  by the superscript
$S$):
\begin{equation}
Q^S(t) = \partial  W(X,t)  ,  \qquad  G_{-}^S=\partial, \quad
Q^S(t,z)=  z \partial  + \partial  W(X,t).
\end{equation}

Here and below the superscript indicating the type of
$QG_{-}$-system will be used only once, i.e. we will
write $H_{Q(t,z)}^{S}$ instead of $H_{Q^S(t,z)}^{S}$.

One can easily show that  $Q^S(t)$  has cohomologies
only  in holomorphic top forms, i.e.  in the $(d,0)$ component,
and  $H_{Q(t)}^S $ is non-canonically  isomorphic to
$J(W)$:
\begin{equation}
H_{Q(t)}^S =   H^S /\{   \omega' dW(t)  \}
\end{equation}
where $\omega'$ is a holomorphic (d-1,0)-form.

The  K.Saito's  $QG_{-}$ system definitely  has no Hodge property,
since the operator $ G_{-}^S$  has no cohomologies.
That  is why relations between different  cohomology  groups
introduced  in subsection 5.1  drastically differ  from those
(Statement  5.2)  that
follow from the Hodge property.

Really,  it is easy to show that  $H_{Q(t,z)}^{S,l}$
has a decreasing filtration (a  class  of  $H_{Q(t,z)}^{S,l}$  is  in
$H(t)^{(k)}$
if  being considered as a class  in  $\hat{H}_{Q(t,z)}^{S}$  it has  a
representative $z^k \omega$):
\begin{equation}
H_{Q(t,z)}^{S,l}=H(t)^{(0)} \supset H(t)^{(1)}  \supset  \ldots
\end{equation}
and
\begin{equation}
0\rightarrow H(t)^{(k+1)} \rightarrow H(t)^{(k)}  \buildrel {\pi_k}  \over
{ \rightarrow} H_{Q(t)}^S \rightarrow 0
\end{equation}
\noindent{\bf  Example.}
For  $d=1$
\begin{equation}
 H(t)^{(k)} =\{   \overbrace{ \partial W(t)\int  ... \partial W(t)
\int }^{k\rm \;  times}
 P(X) dX \} / \{  \partial (W(t)^m),  m \in {\bf
N}    \} .
\end{equation}
Thus, we see that
\begin{equation}
H_{Q(t,z)}^{S,l} \cong  H_{Q(t)}^{S} \otimes C[[z]] \cong
\hat{H}_{Q(t,z)}^{S},
\end{equation}
which is  much larger  than simply  $H_{Q(t)}^{S} $ ,
and the isomorphisms  in the relation above are not canonical.

\noindent{\bf  Definition.}
We define a  "section"  as  a  map  $V(t) :  H_{Q(t)}^{S} \rightarrow
H_{Q(t,z)}^{S,l}$  that  inverts the
projection  $\pi_1$.

Having a "section"  $V(t)$,  we can invert  projections  $\pi_k$ by maps
$z^k V(t)$ ,  thus identifying   $H_{Q(t,z)}^{S,l}  \cong C[[z]]\otimes
Im V(t)$,
and   $H_{Q(t,z)}^{S}  \cong C<<z>> \otimes Im V(t)$.

Finally, the  higher  residue pairings $K_{S}^{(k)}$  are defined  as
a set  of
C-bilinear  pairings  on
$ H_{Q(t,z)}^{S,l}$ .

  It is instructive to  consider  the formal generating  function
\begin{equation}
K_S=\sum_{k\geq 0} z^k K_{S}^{(k)}
\end{equation}
as a pairing
\begin{equation}
K_{S}:  \hat{H}_{Q(t,z)}^{S} \otimes \hat{H}_{Q(t,z)}^{S} \rightarrow  C[[z]].
 \end{equation}

For  original definition see \cite{Sa2};  here, we  will  just
mention some general properties of this pairing  and  present
formulas
for $K^{(0)}$  and   $K^{(1)}$.

\begin{equation}
K_S(z ^k [ \omega_1] ,  z^l  [\omega_2 ] )= (-1)^k  z^{(k+l)} K_S(
[\omega_1] ,
[ \omega_2 ] )
\end{equation}
and
\begin{equation}
K_{S}^{(0)}( [\omega_1] ,  [ \omega_2 ] )=Res  \frac{P_1 P_2
dX^1 \ldots  dX^d }{\prod_{A=1}^{d}  \frac{\partial W}{\partial
X^A}}
\end{equation}
\begin{equation}
K_{S}^{(1)}( [\omega_1] ,  [ \omega_2 ] )=Res  \sum_{A=1}^{d}  \frac{
1/2 (P_2
{\partial \over \partial X^A}P_1 -
P_1 {\partial \over \partial X^A}P_2)
dX^1 \ldots  dX^d }{   \frac{\partial W}{\partial  X^A}
\prod_{B=1}^{d}  \frac{\partial W}{\partial  X^B}}
\end{equation}
 where $[\omega_{\alpha}]\in \hat{H}_{Q(t,z)}^{S}$ for $\alpha=1,2$;
  $P_{\alpha} dX_1  \ldots  dX_d$  is  a  representative of
the class $[\omega_{\alpha}]$ .

\noindent{\bf   Definition.}
K.Saito  defines  the notion  of  a   {\bf   "good section"}
as  a "section"  $V(t)$  satisfying the following  conditions\footnote{
The  action of  $W$  on classes  in (iii) is well defined  since
$W Im QG_{-} \in  Im QG_{-}$,  i.e.
$W\partial \omega \partial W=\partial (W\omega) \partial W$ }:
\begin{itemize}
\item[ (i) ] $ K_{S}^{(k)} (Im V(t), Im  V(t))=0$ , for $ k>0 $
\item[ (ii) ] $\nabla^{GM}  Im  V(t)  \in  z^{-1} Im V(t) + Im V(t)$
\item[ (iii) ] $  [ W(t) Im V(t)]_{Q(t,z)}  \in  Im  V(t)  + z Im V(t) $
\end{itemize}

Using  the notion of a "good section"  K.Saito defines  an improved
connection  $\nabla^S$  on  $\hat{H}_{Q(t,z)}^{S}$  as follows:
when acting on  $H^{(k)}(t)$ for $k>0$,
$\nabla^{S}=\nabla^{GM} $.
For  $[\omega (t)]_{Q(t,z)}  \in   Im V(t)$
\begin{equation}
 \nabla_{i}^{S} [\omega (t)]_{Q(t,z)}=\nabla_{i}^{GM}  [\omega (t)]_{Q(t,z)}
- z^{(-1)}
V( [\Phi_i   \omega(t)]_{Q(t)}).
\end{equation}
and he proves (among  other things) that if  conditions  (i,ii) are
satisfied  his
connection   $\nabla^{S}$  is integrable, and  being restricted  to
$Im V(t)$  preserves the pairing $K_{S}^{(0)}$.

\noindent{\bf   Remark from string theory.} In  "noncompact"
string  theories $\nabla_{i}^{S}$ coincides with the connection
on  the space of all  states (including states, corresponding
to descendents, \cite{Lo2}).

\newsubsec{7.2 The strategy for  reducing  "Landau-Ginzburg"  to K.Saito
theory.}
We can observe the striking  similarity  between K.Saito's connection
$\nabla^{S}$
acting on   $Im V$  and the "Hodge" connection  for  the
"Landau-Ginzburg"
realization  of  "Hodge strings"  system.
Using   this  analogy,  one can guess  that a  "good section"  should
somehow  correspond
to  harmonic  forms.

Naively, K.Saito's  theory  is as far from  theory  with the
Hodge  property as one  can even imagine:
the operator  $G_{-}^S$ has no cohomologies at all.
The second problem is that  K.Saito's theory  is local
while  the global issues seem to be crucial in the harmonic theory.
Below we will overcome these difficulties  as follows.

Here we will  discuss only the  quasihomogeneous case ( $W$ has only one
critical  point) but we expect  that  our methods could be applied  also
for  the general case.
We will consider (in subsection 7.3)  the  "Landau-Ginzburg"
operators  $Q$  and  $G_{-}$ as  operators
acting  on  the space  of
non-holomorphic  germs  of   forms  at  the  critical point  of  $W$.
Below,  we will call  it a  "local LG-system"(omitting the letters
$QG_{-}$ for brevity).
Thus, we have a  natural morphism of  $QG_{-}$ systems,
which induces an isomorphism of  "Gauss-Manin" connections.

 However, the  local  LG-system  does not have the Hodge property
  and  does not
have a  proper pairing.

To study the properties  of
the local  LG-system, we will  introduce the  morphism  $Hol$
that  maps  a non-holomorphic germ to the holomorphic piece of that
germ.
The  morphism $Hol$  maps the local  LG - system to
the K.Saito's  $QG_{-}$  system   and   is   a   quasiisomorphism  of
$QG_{-}$ -
systems.
 Thus,  absence  of the Hodge property  in a local
LG-system  is illustrated  by the  obvious absence
of the  Hodge property in the K.Saito's  system.

Nevertheless,  it is possible  to construct  a "quasihodge"
connection,
(that is   a "pushforward"  of the  "Hodge"  connection
in the global  LG -system), whose covariantly constant  sections  are
germs
  of covariantly
constant sections  of the "Hodge" connection
in the  global  LG-system.

The operation $Hol$  maps the "quasihodge"  covariantly constant
section into  the image  of some  "section" of  K.Saito cohomologies.
We claim (in subsection 7.3) that a "section" obtained in this  way is
a  "good section"
in the K.Saito  sense and show that  it satisfies the condition (ii)
for a  "good section" (see  subsection 7.1).

To  establish  further relations between  local  and
global  LG -systems,  we need   a pairing
on  $\hat{H}_{Q(t,z)}$ cohomologies  in  the  local   system.
The  pairing  (\ref{pair}) is  definitely not
defined  on all germs, so we have to replace it  by  another pairing,
which  should coincide with   (\ref{pair})  on germs of
global  harmonic  forms.  In  doing  this,  we will discover higher
residue pairings  and  their  vanishing  on  germs of  harmonic
forms.
The pairing  we defined  turns  out to be  invariant  under the $Hol$
operation
and gives the  K.Saito  higher  residue  pairings  on holomorphic
germs of
forms. The vanishing  of   higher  residue pairings  on  the
holomorphic pieces   of  germs
of  harmonic forms   mean that they satisfy the condition  (i)
imposed  by  K.Saito  on a "good section"  .

Thus, we  conclude that
(up to condition  (iii))  the image  of  a  "good  sections" in the K.Saito's
 system
is spanned  by   classes  of   holomorphic pieces of  germs  of  harmonic
forms of the global  LG system.

An important issue here is that  K.Saito's  theory  depends
on $\bar{U}$ - but in an obscure way, since  after
passing  to  holomorphic pieces  of  germs,
the
"antiholomorphic
superpotential"  $\bar{U}$  naively  disappears from the problem.
Still  it "shows up" in the choice  of a "good section".
We will discuss this and condition (iii) in the subsection 7.5,
were we find that  condition (iii) is satisfied  if
$\bar{U}$ is quasihomogeneous.

\newsubsec{7.3  Maps  $I$  and $Hol$, and  the condition (ii)
  for a  "good section." }
Suppose  $W(X,0)$  has  an  isolated critical point
at zero.   Let   $H^{g}$   be a space of  non-holomorphic  germs
of forms  at   zero.
There is a natural map  $I: H^F \rightarrow  H^{g}$
given  by expansion  of a form  at  zero.
The operators  $Q(W)$ , $G_{-}(\bar{U})$  and  $\Phi_i$  in the global   LG
system
considered as  operators  on germs lead to
operators  $Q^{g}(W)$ , $G^{g}(\bar{U})$ and  $\Phi_{i}^{g}$
 in the  local  LG system.
Thus, $I$  is a morphism of  $QG_{-}$ systems.

\noindent{\bf  Statement.} The morphism  $I$  induces an  isomorphism
in  $H_{Q(z,t)}$ cohomologies  and
induces  an isomorphism of
"Gauss-Manin"  connections  in these
cohomologies.

What about the Hodge property?
If   $\bar{U}$ also has zero as a critical  point
(and  $dim J(W)=dim J(\bar{U}$)) , then  one can
even  find  simultaneously $Q^{g}(W)$  and  $G^{g}(\bar{U})$
closed  elements in  $H^{g}$ , but   this is  {\it  not  enough} -
the local  LG system  does not have the Hodge property!

To show this  we will   introduce the
operation $Hol$  that  takes  a holomorphic  piece of  a  germ.

\noindent{\bf   Definition.}  We define  the  linear map  $Hol$
from the space  $H^g$  to the  space  $H^S$  of
germs of the  K.Saito   $QG_{-}$  system  as follows:
  $Hol$ sends  $(p,q)$-forms to
zero  if  $q \neq 0$, and
\begin{equation}
Hol (\Omega (X, \bar{X})_{A_1 \ldots  A_p}dX^{A_1} \ldots dX^{A_p})=
\Omega (X, 0)_{A_1 \ldots  A_p}dX^{A_1} \ldots dX^{A_p}.
\end{equation}

It is clear that
\begin{equation}
Hol \circ   Q^g (t,z)  =   Q^S(t,z) \circ  Hol.
\end{equation}
Note that  $Hol$  {\it   wipes  out}    $\bar{U}$.

\noindent{\bf  Statement.}  The map $Hol$  is a quasiisomorphism
between the local  LG-system  and the K.Saito  $QG_{-}$ system.

This  follows from the following Lemma.

\noindent{\bf  Lemma.} For   any  germ  of   holomorphic form  $\omega  \in
H^S$
there is  a germ  $\omega'  \in  H^g$, such that
$\omega + \omega'$  considered  as  an element  of  $H^g$  is
both  $Q^g$ and $G_{-}^g$ closed, i.e.  represents  an
element  of   $H_{Q(t,z)}^{g, l}$.

\noindent{\bf  Idea of the proof.} Below  we will show it  for the case
$d=1$; this
gives the "idea"  why it  happens.

Let us  introduce the parameter $\tau$ in front of  $\bar{U}$,
and   solve  equations  $Q^g(\omega + \omega')=
G_{-}^g(\omega + \omega')=0$,  expanding  $\omega'$  in $\tau$.
Specifically,  let
\begin{equation}\omega'= \sum_{k} \tau^{k} (P_k (X,\bar{X}) dX+ R_k(X,
\bar{X})d\bar{X}),\end{equation}
then  we have to solve the following system:
\begin{equation}
\frac{\partial P_k}{\partial  \bar{X}} -R_k  \frac{\partial
W}{\partial
X}=0;
\end{equation}
\begin{equation}
\frac{\partial \bar{U}}{\partial \bar{X}}P_k - \frac{\partial
R_{k+1}}{\partial  X} =0.
\end{equation}
If  $P_k,R_k$  are known, we can get  $R_{k+1}$ from the second
equation, and then $P_{k+1}$ from the first.
Note that  degrees of  polynomials  $P$ and $R$ are constantly increasing
in these iterations; that is why (in germ topology) the seria
in $\tau$ is convergent.$\Box$

From the previous statement   it follows that
the local LG-system does not
have the  Hodge property (since the K.Saito  system
does not have it).

Nevertheless,  if  $W$ has only one critical  point,
 one can find
a  substitute to the  "Hodge"  connection  in   $H_{Q(t,z)} ^{g}$;  let
us
call it the "quasihodge" connection.

\noindent{\bf   Definition.} Let   $\omega_{a}^{H}(t)$  be
harmonic elements in $H^F$ that are covariantly constant
with respect to "Hodge" connection.
Let  us define  "quasiharmonic"  elements  in $H^g$
as germs  of  expansion  at  zero  of  harmonic  elements
in $H^F$:  $\omega_{a}^{QH}(t)= I  \omega_{a}^{H}(t)$.
"Quasiharmonic" elements   determine
 the  "quasihodge"  connection  $\nabla^{QH}$ in $H_{Q(t,z)}^{g}$
through  its  covariantly constant  sections
$[P_{a}(z,z^{-1}) \omega_{a}^{QH}(t)] _{Q^{g}(t,z)}$, where $P$ are
polynomials.

  "Quasiharmonic"  elements  satisfy equation
(that  is obtained  by expansion at zero of the corresponding
equation in global LG system, see subsection 5.2 ):
\begin{equation} \label{qh}
\frac{\partial}{ \partial  t_i}  \omega_{a}^{QH}(t) + z^{-1}( \Phi_{i}^{g}
\omega_{a}^{QH}(t) -
C_{i,a}^{b}(t) \omega_{b}^{QH}(t)) \in  Im Q^{g}(t,z)
\end{equation}

 Relation  between  "quasihodge"  and "Gauss-Manin"
connections  is exactly the same as in global LG-system:
\begin{equation}
\nabla_{i}^{GM} =
\nabla_{i}^{QH} - z^{-1} C_{i}.
\end{equation}

Let  us define the "quasiharmonic"  elements
 $\omega_{a}^{QHS}(t) \in H^S$
as images of  $Hol$  acting on  "quasiharmonic" elements  in
$H^g$:
\begin{equation}
 \omega_{a}^{QHS}(t)=Hol  \omega_{a}^{QH}(t)= Hol  \circ  I  \omega_{a}^{H}(t)
 \end{equation}
As in local  LG-system "quasiharmonic" elements determine the
"quasihodge" connection  $\nabla^{QHS}$  in $H_{Q(t,z)}^S$ that
is related to the "Gauss-Manin" connection  $\nabla^{GMS}$
like in local  LG-system (just apply $Hol$ to (\ref{qh})):
\begin{equation} \label{qhs}
\nabla_{i}^{GMS} =
\nabla_{i}^{QHS} - z^{-1} C_{i}.
\end{equation}

Classes  of  "quasiharmonic" elements  in  $H_{Q(t,z)}^S$
span the  vector  space  $H^{QHS}(t)  \subset  H_{Q(t,z)}^S$
that  projects onto  the space $H_{Q(t)}^S$, i.e.
the space $H^{QHS}(t)$  could be  considered  as
the image  of  the  "section" $V^{QHS}(t)$:
\begin{equation}
H^{QHS}(t)=Im V^{QHS}(t)
\end{equation}
From  (\ref{qhs})  we conclude that
the  "section"  $V^{QHS}(t)$ satisfies the
K.Saito's  condition (ii)
for a  "good  section" (over the space $C[[t]]$).

Now we need  a pairing.

\newsubsec{7.4 Higher pairings.}
In this  subsection we continue to assume that  $W(X,0)$ has
only one critical point at $X=0$.
We  will try  to define a pairing  on   $H_{Q(t,z)}^{g,l}$
that  coincides  with  the pairing  (\ref{pair})  on  germs of forms from
$H^F$.

\noindent{\bf  Definition.}
Consider the space  $H^P$ of forms on $C^d$ whose coefficients
grow not faster than a polynomial as $|X| \rightarrow  +\infty$.
Let us take an operator  $R=*\bar{\partial}\bar{W}* $, where $*$ is a
Hodge
operation. Take a positive real  number $\epsilon$ .
Then, we define the bilinear pairing $<,>(\epsilon, z)$  on  $H^P$
with values in  $C[[z]]$  as:
\be
<\omega_1,\omega_2>_P(\epsilon,z)=\int_{C^d} \omega_{1} {\cal C}
\exp (- \epsilon^{-1} \{  Q(t)+zG_{-} ,  R \} )
\omega_2
\ee
Here ${\cal  C}$ is a Weil  operator (see (\ref{pair})).

\noindent{\bf   Statement  7.4.1}
The  asymptotic  expansion
at  $\epsilon =0 $   of  the pairing  $<,>_P(\epsilon, z)$ leads to
a  pairing  $<,>_g(\epsilon, z)$  defined  on germs at  $X=0$.

\noindent{\bf  Proof.}  Fix the  power   $n$  of   $z$, and   then
use the saddle
point
estimation
of the  pairing on  forms  whose coefficients are  monomials.
The $\epsilon$ expansion  of the pairing  has the  form:
$(\epsilon)^{(m-k)/L}  S(\epsilon^{1/L})$,   where $k$ and $L$  are
some
integers
depending on the polynomial  $W$ and the integer $n$, the positive integer
$m$
depends
on the monomial,   and  $S(u)$ is some Taylor seria.  As the power  of the
monomial
grows,  $m$  tends to infinity.$\Box$

\noindent{\bf  Definition-Statement  7.4.2}
The value  of the pairing  $<\omega_1 , \omega_2>_g(\epsilon, z)$
on  $Q(t)$ and $G_{-}$  closed  germs  is  independent  of
$\epsilon$ and defines  the pairing  $<,>_{g,\bar{U}}(z)$  between
$Q(t)-zG_{-}$
and  $Q(t)+zG_{-}$ cohomologies in the space of germs.

\noindent{\bf  Proof.} Take the derivative in $\epsilon$. This brings down
the  $Q(t)+zG_{-}$  exact term.  After  commutation with ${\cal C}$, the
operator
$Q(t)+zG_{-}$  turns into  $Q(t)-zG_{-}$. Thus, the derivative in $\epsilon$
is equal  to zero. Similarly one can prove  that  the change
of  $\omega_2$ by  $Im (Q(t)+zG_{-})$ or of  $\omega_1$
by  $Im (Q(t)-zG_{-})$  does not change the value of  the pairing.$\Box$

\noindent{\bf  Statement  7.4.3 }
Main property of   the pairing  $<,>_{g,\bar{U}}( z)$ :
the value of this pairing on germs of harmonic forms
from $H^F$ is independent  on    $z$  and  coincides  with
the value of the pairing (\ref{pair}) on
corresponding harmonic forms:
\begin{equation}
<I \omega_{a}^{H}, I \omega_{b}^{H} >_{g,\bar{U}}(z)=
< \omega_{a}^{H},  \omega_{b}^{H} >
\end{equation}

\noindent{\bf  Proof.}  Consider  the value of
the pairing  $<,>_P(\epsilon, z)$   on  harmonic
forms  from  $H^F$  as  a smooth  function  of  $\epsilon$,
taking values in $C[[z]]$.
As  in the proof of  Statement 7.4.2  we conclude  that such  a function
is  independent  of  $\epsilon$.  In order to  evaluate this function
we
can go to  the limit
$\epsilon \rightarrow  +\infty$. Such  a limit  exists and is equal
to  the
pairing $<,>$ on  the harmonic forms  from  $H^F$ ; so, the  $z$ dependence
disappears.$\Box$

\noindent{\bf  Definition.} Let us define the pairing
$<,>_{Hol}(z)$  on  holomorphic germs of  $(d,0)$-forms as
a  pairing  $<,>_{g,0}(z)$.
Expanding in  $z$,  we get a set of
higher  pairings  $<,>_{Hol}(z)= \sum_{k} <,>_{Hol}^{(k)} z^{k}$

\noindent{\bf  Statement  7.4.4}

\begin{equation}
<\omega_1, \omega_2  >_{g,\bar{U}} (z)=
 <Hol  (\omega_1),  Hol  ( \omega_2)> _{Hol} (z).
\end{equation}

\noindent{\bf  Proof.}  Consider  the antiholomorphic dilatation
$D_{\lambda}$:
 $X \rightarrow X$,  $\bar{X} \rightarrow \lambda  \bar{X}$.
The dilatation  leaves  $Q(t)$ invariant  but  transforms
$G_{-} (\bar{U})$ into
$G_{-}(D_{\lambda} \bar{U})$ .
It also  transforms   $ \{  Q(t)+zG_{-}(\bar{U}) ,  R \} $ to
 $ \{  Q(t)+zG_{-}(D_{\lambda} \bar{U}) , D_{\lambda} R \} $.
The  change  in   $R$ does not change  the
 value of  the pairing on  $Q(t)$  and $G_{-}$
closed  germs of forms( if this value is defined)
by  the argument  of "bringing  down"
a  $Q(t,z)$ exact term from the exponent;  thus,
 \begin{equation}
<\omega_1, \omega_2  >_{g,\bar{U}} (z)=
<D_{\lambda}\omega_1, D_{\lambda}\omega_2  >_{g,
D_{\lambda}\bar{U}} (z).
\end{equation}
Now,  taking $\lambda$ to zero,  we  have proved  the statement.$\Box$

\noindent{\bf   Conjecture.}
The K.Saito pairing  $K^S(z)$  coincides  with  $<,>_{Hol}(z)$.

Arguments in favor of  the  conjecture:\\
1) Both pairings are in some sense natural
in   $Q(t,z)$  cohomologies.\\
2)By explicit computations  one can show that
the first two  terms in the expansion  in $z$ coincide for
these  two pairings.\\
3)The  $(-1)^k$   factor in K.Saito's pairing  could be
explained  by observing that,  after commutation
with the  Weil   operator  ${\cal  C}$,
$Q(t)+zG_{-}$  passes to $Q(t)-zG_{-}$.$\Box$

Putting everything together,  we see that  higher pairings
$<,>_{Hol}^{(k)} $, $k>0$  vanish  on  holomorphic pieces
of  germs of  covariantly constant  sections  of the "Hodge" connection:
\begin{equation}
< \omega^{QHS} ,\omega^{QHS}
>_{Hol}^{(k)}
=
< I( \omega^{H}) ,  I  (\omega^{H})
>_{g,\bar{U}}^{(k)} =0  , k>0.
  \end{equation}

Thus the "section"  determined by  $V^{QHS}$ (i.e.  whose image is spanned
by classes  of  holomorphic pieces  of germs of harmonic forms)
satisfies  the  condition  (i) of K.Saito for
a "good section"  (if we assume  that conjecture above is correct).

\newsubsec{7.5   "Holomorphic anomaly"  and  K.Saito's condition (iii) }
 Here we continue to assume that  $W(X,0)$ has
only  one critical point at $X=0$.

From results of subsections (7.3) and  (7.4) we see that
taking  any  $\bar{U}$ that is good  for  $W$  we can
construct  $V^{QHS}(t, \bar{U})$ that determines  a "section",
that satisfies  conditions (i)  and (ii).
Now we will study how  $V^{QHS}(t,\bar{U})$ depends on $\bar{U}$.

Consider  the family  inside the space of
polynomials  $\bar{U}$ that are good for $W(X,t)$:
\begin{equation}
\bar{U}(\bar{X}, t')=
  \bar{U} (\bar{X})+ t'\Psi(\bar{X})
\end{equation}
where  $\Psi (\bar{X})$ is a polynomial, depending on
$\bar{X}$,  $T'$  is a base of the family  and  parameter
$t'$ is a  coordinate  on the base.

 Harmonic elements
of the pair  $(Q(W(t)), G_{-}(\bar{U}(t'))$ form a bundle over
$T'$.
There  is  a  "Hodge'  "  connection in this bundle
such that  the  classes  in  $Q(t)$-cohomologies
of its  covariantly constant  sections do not depend on  $t'$.
We denote these covariantly  constant sections as  $\omega_{a}^{H'}(t',t)$,
where index  $a$ labels  some basis in  $Q(t)$-cohomologies.

Operator $Q(t)+zG_{-}(t')$ we will denote as $Q(t,t',z)$.

Let us  define  classes  $[\omega_{a}^{QH'S}(t',t)]_{Q^S(t,z)}$ as:
\begin{equation}
[\omega_{a}^{QH'S}(t',t)]_{Q^S(t,z)}=[Hol \circ I
\omega_{a}^{H'}(t',t)]_{Q^S(t,z)}
 \end{equation}
 Note, that
    $H^{QHS}(t,\bar{U}(t') )=
Span \{ [\omega_{a}^{QH'S}(t',t)]_{Q^S(t,z)}  \}$  .

Since polynomial  $\Psi$ ( considered as an operator acting
in the space of forms by multiplication) commutes with $G_{-}(t')$,
this polynomial leads to a linear
operator  acting in the space of
$G_{-}(t')$ cohomologies.
 This action is represented by
a linear operator  $\bar{C}(\Psi, t',t)$  whose matrix elements
in the basis $[\omega_{a}^{H'}(t',t)]_{G_{-}(t')}$ we will denote
 as  $\bar{C}(\Psi, t',t)_{a}^{b} $:
\begin{equation}
\Psi \omega_{a}^{H'}(t,t')=C(\Psi,t',t)_{a}^{b}
\omega_{b}^{H'}(t,t') + Im G_{-}(t')
\end{equation}

\noindent{\bf  Statement   7.5.1} In $Q^S(t,z)$  cohomologies
\begin{equation}
\frac{\partial}{\partial  t'} [\omega_{a}^{QH'S}(t',t)]_{Q^{S}(t,z)}=
z(\bar{C}(\Psi,t',t)_{a}^{b}- \Psi(0)\delta_{a}^{b})
[\omega_{b}^{QH'S}(t',t)]_{Q^{S}(t,z)}
 \end{equation}

\noindent{\bf  Proof.}
By reasoning  like  in the proof  of  step 1  in  subsection 5.1
we get
\be
 \frac{\partial}{\partial  t'} \omega_{a}^{H'}(t',t) +
z (\Psi(\bar{X}) \omega_{a}^{H'}(t',t) -
\bar{C}(\Psi)_{a}^{b}(t',t)\omega_{b}^{H'}(t',t)) \in Im(Q(t,t',z))
 \ee
After passing to germs and
taking the  holomorphic piece we are left  only  with
$\Psi(0)$. $\Box$

\noindent{\bf  Remark.} A version of  statement  7.5.1
 is known in the physics literature as $t-t^*$ equations \cite{CV}.

Up to now  we ignored  condition (iii) of  K.Saito.
Now we are in position to study  restrictions it imposes
on  $\bar{U}$.

\noindent{\bf  Statement   7.5.2}
Let  $u=\bar{U}(0,t')$.  Then,
\begin{eqnarray}
&&[W(t) \omega_{a}^{QHS}(t,t')]_{Q^S(t,z)}  \in \nonumber \\
&&   H^{QHS}(t,t') +
z  H^{QHS}(t,t') +z^2  (\bar{C}(\bar{U}(t'))- u  ) H^{QHS}(t,t')
\end{eqnarray}

\noindent{\bf  Proof.}  Let  us fix $t$ and $t'$.
Consider auxiliary  family $T''$ of pairs
of polynomials $\bar{U}(t',t'')=(1+t'')\bar{U}$,
  $W(t, t'')=(1+t'')^{-1}W(t)$; $t''$ is a coordinate
in $T''$.   Now take  two
connections ($\nabla^A$ and $\nabla^B$)  in the bundle of harmonic  forms
over   $T''$. First we describe connection $\nabla^A$.

Consider the manifold of "good pairs" of polynomials
$W$ and $\bar{U}$, and a bundle of harmonic forms over this manifold.
There is a "sum" Hodge connection in this bundle, that is defined as follows.
The restriction of the "sum" Hodge connection to submanifolds of varying
$W$ for fixed $\bar{U}$ is a Hodge connection (see section 5.2),
and its restriction to  submanifolds of varying $\bar{U}$ for fixed $W$
is a Hodge' connection that was already defined in this subsection.
The family $T''$ is a submanifold in the space of "good pairs"
and connection $\nabla^A$ is a restriction of the "sum"
Hodge connection to $T''$.

 Connection $\nabla^B$ follows  from (\ref{dil})  and  is induced by
multiplication with $(1+t'')^{\hat{q}}$, where  $\hat{q}$ is an operator
acting  on a $(p,q)$-form as multiplication by $q$.  From  comparison of the
two connections at $t''=0$ we get:  \begin{equation} (\nabla^A-\nabla^B)
 (Span\{ \omega_{a}^{H'}(t,t') \}) \subset (Span\{ \omega_{a}^{H'}(t,t')\} )
 \end{equation}
 Evaluation  of  this  relation up to the image of $Q(t,t',z)$ gives:
\begin{eqnarray}
&& z (\bar{U}(\bar{X})
\omega_{a}^{H}(t,t') - \bar{C}(\bar{U})_{a}^{b}\omega_{b}^{H}(t,t')) -
 \nonumber \\
&& (z^{-1}) (W(X,t) \omega_{a}^{H}(t,t') -
 C(W(t))_{a}^{b}\omega_{b}^{H}(t,t'))- \nonumber \\ && - \hat{q}
 \omega_{a}^{H}(t,t') \in Span\{ \omega_{c}^{H}  \} +ImQ(t,t',z)
 \end{eqnarray}
 Now
application of  $Hol \circ  I$ proves the statement.$\Box$.

\noindent{\bf Corollary.}
 K.Saito condition
(iii) ( for a case with  one critical point)  is satisfied if  $\bar{U}$ is
quasihomogeneous.

\noindent{\bf Proof.} If $\bar{U}$ is quasihomogeneous,
 $\bar{U}$ is $G_{-}$ exact and acts by zero in $G_{-}$
cohomologies, i.e.
$\bar{C}(\bar{U})_{a}^{b}=0$. Then the condition (iii) of
K.Saito follows from statement 7.5.2.$\Box$

Note, that condition (iii) of  K.Saito leaves open a possibility to
get  different "good sections" from different quasihomogeneous $\bar{U}$
(phenomena of "holomorphic anomaly").

\noindent{\bf  Example of holomorphic anomaly.}
Consider  $W(X,Y)=X^4+ Y^4$,  and
 \begin{equation}\bar{U}(\bar{X},
\bar{Y},t')=\bar{X}^4+ \bar{Y}^4+
t' \bar{X}^2  \bar{Y}^2.
\end{equation}
Here "good section"   $V^{QHS}(t')$ does  depend on
$t'$. In particular,
\begin{equation}  \label{anom}
V([X^2Y^2 dXdY]_{Q^S};t')=[X^2Y^2 dXdY + zc(t')dXdY ]_{Q^S(z)},
\end{equation}
with some function $c(t')$   that is  not zero (but
$c(0)=0$). Really, $\bar{X}^2\bar{Y}^2$ is non-zero
in $G_{-}$ cohomologies, and statement of holomorphic
anomaly follows from the Statement 7.5.1.
One can also directly check that the family
of sections (\ref{anom}) do satisfy (i) and (iii) requirements of
K.Saito.

\section{8. Conclusion}
In this paper  we have shown  that  genus zero "amplitudes"  in
"compact" topological string  theories are completely determined
by  the corresponding "Hodge string"  $QG_{-}$ system.

It seems that  $QG_{-}$ systems deserve the study on their own.
They  form a  category  similar  to
the  category  of manifolds.
"Hodge string"  $QG_{-}$ systems (like  global  "Landau-Ginzburg"
system) look like compact  manifolds,
i.e.  there are theorems of existence of different  structures but
it is hard to compute them.
The $QG_{-}$ systems of the  K.Saito type are like
affine manifolds , and  the  choice of  "good section"
resembles compactification.

\vspace{1cm}

\centerline{\bf Appendix: From the family of  hypersurfaces}
\centerline{\bf  to $z\partial+
\partial W(t)$ cohomologies.}

\vspace{1cm}

In this appendix  we will  show (following  K.Saito) how  to get from
 the family of  hypersurfaces  to the bundle  of
$z\partial +  \partial W(t)$ cohomologies, and how the
Gauss-Manin  connection  in the cohomologies of the
fiber  in K.Saito  interpretation
leads to canonical  "Gauss-Manin" connection (subsection (5.1)).

In this Appendix  $\partial$ stands for
partial derivatives in $X$ only.

Consider  the family of  affine hypersurfaces in $C^d$
over  $S\otimes T$  defined
by equation :
\begin{equation}
W(X,t)-s=W(X)+t_i \Phi_i(X)-s=0.
\end{equation}
Here, parameters  of equation  $s$, $t_i$  are
considered as  algebraic coordinates on $S$, and
$T$ respectively.
This  surface  is degenerate  on the discriminant  $Dis$,
defined as  the submanifold  in $S\otimes T$ , such that
\begin{equation}
(W(t)-s) \in  C[X]/ I(W),
\end{equation}
where $I(W)$ is the ideal generated by partial derivatives
of $W$ with respect to $X$.
For    $(s,t)  \in S\otimes T- Dis$  the  surface is nondegenerate,
and the holomorphic (d-1) forms on it  are given
as  residues  of the  meromorphic  forms
\begin{equation}
\frac{\Omega(t,s)}{W(t)-s},
\end{equation}
where $\Omega$ is a holomorphic $d$ form on $C^d$.
If
\begin{equation}
\Omega=\partial W(t) \partial \omega,
\end{equation}
then,  its  periods vanish (just integrate $\partial \omega$ by
parts);
so, it is exact.
Thus,  $(d-1,0)$-cohomologies  $H(s,t)$
of the surface  over  $S\otimes T - Dis$ are equal to:
\begin{equation}
H^{(0)}(s,t)=\{ \Omega( s,t )  \}/  \{  (W(t)-s) \Omega(s) +
\partial
\omega(s) \partial W \}.
\end{equation}

If  $(s,t)  \in  S \otimes T - Dis$,
there is another  way  to represent  $(d-1,0)$-forms
 on the hypersurface as a residue:

\begin{equation}
\frac{\omega(s,t) \partial  W(t)}{W(t)-s}.
 \end{equation}

Thus, we can define
 \footnote{
 for $d=1$   $\partial \omega'(s,t)$ has to be replaced
by  a  $X$-independent function of $s$ and $t$ because $\partial$ has
cohomology in the space of  holomorphic 0-forms}

\begin{eqnarray}
&&H^{(1)}(s,t)= \\  \nonumber
&&\{ \omega(s,t) \partial W(t) \}/
 \{ ( (W(t)-s) \Omega(s,t) + \partial \omega'(s,t)
 \partial W(t) ) \cap \omega(s,t) \partial W(t)  \} ,
\end{eqnarray}
and  there  is   a  map
 \begin{equation} b(s,t) :    H^{(1)}(s,t) \rightarrow  H^{(0)}(s,t)  ,
 [\omega(s,t) \partial W(t)]_{(1)} \mapsto  [\omega(s,t) \partial W(t)]_{(0)}
\end{equation}
 that   is an  isomorphism  for  $(s,t)  \in  S \otimes T - Dis$.
Here,  $[ \;]_{(i)}$ mean  equivalence  classes  in $H^{(i)}$.

There is a Gauss-Manin  connection, defined as a unique
flat  connection  in the  bundle  of   $(d -1,0)$ cohomologies of
the family of the non singular  hypersurfaces, such that  the periods
of
its covariantly constant sections  are constant  (as functions of
parameters).

  This connection acting on cohomologies
(written in the form $H^{(0)}(s,t)$)   has the following  form:
\begin{equation}
\nabla_{s}^{GM} [\Omega(s,t)]_{(0)} =[\partial  \omega(s,t)]_{(0)}
\end{equation}
\begin{equation}
\nabla_{i}^{GM} [\Omega(s,t)]_{(0)} =[\Phi_i \partial  \omega(s,t)]_{(0)}
\end{equation}
for   $[\omega(s,t)  \partial W(t)]_{(1)}=b^{(-1)}(s,t)([
\Omega(s,t)  ]_{(0)}).$

The  definitions  of   $H^{(i)}$  could be easily extended
to $(s,t) \in  Dis $, while (as one can expect)
for such values of  $(s,t)$,  the morphism  $b(s,t)$  is just
an inclusion, and  (as one could expect from the very beginning)
  the Gauss-Manin
connection is not  defined.
It becomes a  Gauss-Manin operator
(that  we will  still denote  by the same letter
$\nabla^{GM}$ ) acting  from  $H^{(1)}(s,t)$  to  $H^{(0)}(s,t)$
considered as modules over  ${\cal  O}(S \otimes T)$ .
K.Saito  introduced  the decreasing
filtration  on the module  $H^{(0)}(s,t)$:
class $[\Omega]_{(0)} \in  H^{(k)}$  iff
$(\nabla_{s}^{GM})^k [\Omega]_{(0)} $  is
well defined.

In order  to have  Gauss-Manin connection  defined
everywhere K.Saito  extends  the space  $H^{(0)}$ to
the space
\begin{equation}
H(\delta_{s}^{-1}) =
C<< \delta_{s}^{-1}>> \otimes_{C[\delta_{s}^{-1}]} H(s,t)
\end{equation}
and  $\delta_{s}^{-1}$ acts on $H(s,t)$ as an inverse of
the Gauss-Manin connection $\nabla_{s}^{GM}$.

Now the Gauss-Manin connection  acts  on  the space
$H(\delta_{s}^{-1})$  as follows:
\begin{equation}
\nabla_{s}^{GM}   \delta_{s}^{-k}[\omega]_{(0)} =
\delta_{s}^{-(k-1)}[\omega]_{(0)},
\nabla_{i}^{GM}   \delta_{s}^{-k}[\omega]_{(0)} =
\delta_{s}^{-(k-1)}[  \Phi_i  \omega]_{(0)}
\end{equation}

In order to make contact  with  the subsection 7.1  we first  observe
that  over  $C[s]$  the space $H^{(0)}$ could  be  identified
with
\begin{equation}
\{ \Omega(t) \} /\{\partial W(t) \partial \omega(t) \}
\end{equation}
$P(X,t,s)dX^1\ldots  dX^d \mapsto P(X,t,W(X))dX^1\ldots  dX^d $.
It  is still a ${\cal O}[s]$ module, with  $s$ acting as multiplication with
$W(X)$.

Recalling that the space of  holomorphic top forms  was named  in
subsection 7.1 as   $H^S$,  and that  operators  $\partial$ and  $\partial
W(t)$
were named  as  $G_{-}^S$ and  $Q^S(t)$  respectively, we  identify
\begin{equation}
H^{(0)} \cong H^S/ Im Q^S(t)G_{-}^S
\end{equation}

i.e.  $H^{(0)} \cong H^{l,S}$.
One can see that  K.Saito's  filtration (described in this Appendix)
corresponds to filtration by powers of  $z$ that is induced on
$H^{l,S}$ from  its inclusion  in  $\hat{H}_{Q(t,z)}$ (see section 5.1).

Thus, identifying  $\delta_{s}^{-1}$  with  a  formal parameter $-z$
from subsection 5.1,
one easily finds  that  K.Saito's   $H(\delta_{s}^{-1})$
   corresponds to $H_{Q(t,z)}^S$
and  Gauss-Manin connection  acting  $H(\delta_{s}^{-1})$
 corresponds to canonical
"Gauss-Manin" connection (see subsection (5.1)) .

\vskip 1pc
\noindent{\it Acknowledgments.}
I  am grateful  to  A. Gerasimov
for sharing  his ideas  about relations between
$QG_{-}$ systems , strings  and  Hodge theory.
I would like to thank  R.Dijkgraaf  for  comments  on relation
between "Hodge"  connection and  CPA connection,
N.Nekrasov for  explaining  to me relations between  K.Saito's
filtration and  asymptotics of  oscillating integrals,
K.Saito  for explaining to me his theory and
A.Varchenko  for  insisting  that  theory  has to be formulated
in terms of germs.
I would like to thank  V.Batyrev, A.Beilinson,  A.Givental,
 G.Moore, I.Polyubin, A.Rosly  and
S.Shatashvili  for helpful discussions.
I am grateful to organizers of Taniguchi Symposium
for providing an excellent conditions for exchange of
scientific ideas.

\def\thebibliography#1{\vskip 1.2pc{\centerline {\bf References}}\vskip 4pt
\list
 {[\arabic{enumi}]}{\settowidth\labelwidth{[#1]}\leftmargin\labelwidth
 \advance\leftmargin\labelsep
 \usecounter{enumi}}
 \def\newblock{\hskip .11em plus .33em minus .07em}
 \sloppy\clubpenalty4000\widowpenalty4000
 \sfcode`\.=1000\relax}
\let\endthebibliography=\endlist

%\begin{thebibliography}{ABCD}

\vskip 2pc
\noindent Andrei S. Losev\\
\noindent Institute of Theoretical and Experimental Physics ITEP\\
\noindent 117259,B.Cheremushkinskaya,Moscow,Russia ,  and \\
\noindent   Department of Physics,Yale University,New Haven,CT 06520\\
\noindent email:lossev@vitep3.itep.ru, lossev@waldzell.physics.yale.edu
\vskip 6pt

\end{document}